\documentclass[aip, jcp, reprint]{revtex4-2}
\usepackage{amsmath,amsthm,amssymb,amsfonts,bm,verbatim,graphicx,multirow}
\usepackage{array,graphics,rotating,cancel,color}
\usepackage[mathscr]{euscript}
\usepackage{hyperref}
\usepackage{threeparttable}

\usepackage{hyperref}
\hypersetup{colorlinks,breaklinks,linkcolor=Blue,urlcolor=Blue,anchorcolor=Blue,citecolor=Blue}
\usepackage{color}
\definecolor{DarkBlue}{rgb}{0.0,0.08,0.45}
\definecolor{Blue}{rgb}{0.0,0.0,1.0}
\definecolor{Red}{rgb}{1.0,0.0,0.0}
\definecolor{RedOrange}{rgb}{0.9,0.0,0.2}
\definecolor{dgrn}{RGB}{0,150,0}
\definecolor{dgray}{gray}{0.3}

\usepackage{dcolumn}

\newcommand{\ie}{{\em i.e.}}
\newcommand{\eg}{{\em e.g.}}
\newcommand{\etal}{{\em et al.}}

\newcommand{\mc}[3]{\multicolumn{#1}{#2}{#3}}
\newcommand{\fns}{\footnotesize}
\newcommand{\fig}[1]{\scalebox{0.6666}{\includegraphics{figure/#1}}}
\newcommand{\code}[2]{\scalebox{#1}{\includegraphics{code/#2}}}

\newcommand{\pref}[1]{\protect\ref{#1}}

\newcommand{\jax}{\textsc{Jax}}

\newcommand{\pyscf}{\textsc{PySCF}}
\newcommand{\pyscfad}{\textsc{PySCFAD}}
\newcommand{\libcint}{\textsc{Libcint}}
\newcommand{\libxc}{\textsc{Libxc}}

\newcommand{\numpy}{\textsc{NumPy}}
\newcommand{\scipy}{\textsc{SciPy}}
\newcommand{\qchem}{\textsc{Q-Chem}}

\begin{document}
\title{
    Differentiable quantum chemistry with {\pyscf} for molecules and materials at the mean-field level and beyond
}

\author{Xing Zhang}
\affiliation{Division of Chemistry and Chemical Engineering, California Institute of Technology, Pasadena, CA 91125}
\author{Garnet Kin-Lic Chan\footnote{{\tt gkc1000@gmail.com}}
}
\affiliation{Division of Chemistry and Chemical Engineering, California Institute of Technology, Pasadena, CA 91125}
\date{\today}

\begin{abstract}
    \noindent
    We introduce an extension to the {\pyscf} package which makes it automatically differentiable.
    The implementation strategy is discussed, and example applications are presented to demonstrate the
    automatic differentiation framework for
    quantum chemistry methodology development.
    These include orbital optimization, properties, excited-state energies, and derivative couplings, at the mean-field level and beyond, in both molecules and solids. We also discuss some current limitations and directions for future work.
\end{abstract}

\maketitle

\section{Introduction}
Automatic differentiation (AD)~\cite{Wengert1964,Griewank2008}
has recently become ubiquitous through its adoption in machine learning applications.
AD removes the burden of implementing  derivatives of
complex computations by applying the chain rule to
the elementary operations and functions of a computational graph.
This can be used to automatically compute exact derivatives of arbitrary order.

In quantum chemistry, derivative evaluations appear in many types of calculations, such as in wavefunction optimization,
to compute response properties, to obtain critical points and reaction paths on potential energy surfaces,
in molecular dynamics simulations, for basis set and pseudopotential optimization, {\em etc}.
Before the advent of AD, these derivatives were mainly computed by analytic schemes, where the symbolic formula for the derivative is first required (and which can be tedious to obtain), or computed numerically using finite difference methods (which are computationally expensive and
prone to numerical instabilities).
Both sets of difficulties are circumvented by applying AD, as shown in a few works very recently.
For example, Tamayo-Mendoza {\etal}\cite{Tamayo-Mendoza2018}
implemented a fully differentiable Hartree--Fock (HF) method with AD;
Song {\etal}\cite{Song2020} introduced an AD scheme
to compute nuclear gradients for tensor hyper-contraction based methods;
Abbott {\etal}\cite{Abbott2021} applied AD to calculations of higher order nuclear derivatives
with methods such as HF, second-order M{\o}ller--Plesset perturbation theory (MP2)
and coupled cluster theory with single, double and perturbative triple excitations [CCSD(T)]; and
Kasim {\etal}\cite{Kasim2022} developed a differentiable quantum chemistry code called DQC
for basis set optimizations, molecular property calculations {\em etc.} at the mean-field level.
In addition, there are other quantum chemistry related applications where AD has played an important role,
{\eg}, in training neural network based density functionals,\cite{Li2021,Kasim2021}
in differentiable tensor network algorithms,\cite{Liao2019}
and in quantum circuit optimization.\cite{Arrazola2021}

A central mission of {\pyscf}~\cite{PySCF1,PySCF2}  is to provide a development platform to accelerate the implementation of new methods by its users.
To further its potential, we have attempted to incorporate into {\pyscf} the full functionality of AD.
The resulting AD framework, which we call {\pyscfad} in the following,
is available as an add-on to {\pyscf}.\cite{PySCFAD}
Although still in active development, we have found that {\pyscfad} is already
very useful for derivative calculations associated with complex computational workflows. The purpose of this paper is thus to describe the current implementation of AD within {\pyscfad} and to illustrate its use in a range of applications.

The remainder of the paper is organized as follows.
In section \pref{sec:imp}, we discuss the implementation of a few key components
in {\pyscfad}. In section \pref{sec:app}, example applications are presented to illustrate the
capabilities of {\pyscfad}.
In section \pref{sec:speed}, we examine the computational efficiency of {\pyscfad}.
In section \pref{sec:conc}, we summarize our experience with AD in this project, and outline some directions for future work.

\section{Implementation} \label{sec:imp}
We now describe the implementation of {\pyscfad}.
We will assume some knowledge of the basics of AD; for further introduction see Ref.~\cite{Griewank2008}.
In {\pyscf}, a majority of methods are implemented using
{\numpy}\cite{Numpy} and {\scipy}\cite{Scipy} functions.
These functions have differentiable counterparts
provided by several advanced AD libraries.\cite{Walter2013,Paszke2017,jax2018github}
(Here differentiable means that the function (or data) can be used or traced  by an AD library in the computation of derivatives).
Thus in many scenarios, transforming {\pyscf} into {\pyscfad}
is simply a matter of replacing the {\numpy} and {\scipy} functions
with the corresponding differentiable ones from the AD library.
However, there are also computationally heavy components such as the electron
repulsion integrals (ERIs) which are implemented as C code in {\pyscf}.
We realize AD for these parts by registering the C functions that implement their analytic derivatives, which allows them to be called during the AD traversal of the computational graph.
This not only avoids duplicate implementations of the same functionality,
but also reduces the cost of AD tracing through complex numerical algorithms.
Note, however, that
such a strategy only allows for derivatives of finite order,
as the C derivative functions appear as black-boxes to the AD framework.

Currently, we use {\jax}\cite{jax2018github} as the backend AD package for {\pyscfad}.
This is because besides being an AD library, {\jax} also provides other appealing features such as
vectorization, parallelization, and just-in-time compilation, including for hardware accelerators.
Using the strategy above, the following components of {\pyscf} have been transformed to be compatible with automatic differentiation:
\begin{itemize}
\item Molecular and crystal structures, Gaussian orbital evaluations and ERIs ({\em gto} and {\em pbc/gto})
\item Molecular and plane wave density fitting routines ({\em df} and {\em pbc/df})
\item HF and density functional theory (DFT) for molecules and solids
({\em scf}, {\em dft}, {\em pbc/scf} and {\em pbc/dft})
\item Time-dependent HF and DFT ({\em tdscf})
\item MP2 ({\em mp})
\item Random phase approximation ({\em gw})
\item Coupled cluster theory ({\em cc})
\item Full configuration interaction ({\em fci}).
\end{itemize}
(The relevant modules in {\pyscf} are listed in parentheses.)
In the following, we give more details about the implementations of some of them.

\subsection{Electron repulsion integrals}
{\pyscf} uses contracted Gaussian basis functions, the radial parts of which can be expressed as
\begin{equation}
    \phi(\mathbf{r}) = \sum_i C_i (\mathbf{r}-\mathbf{r}_0)^l \exp(-\alpha_i (\mathbf{r}-\mathbf{r}_0)^2) \;,
\end{equation}
where $l$ is the angular momentum, and the three parameters
$\mathbf{r_0}$,  $\alpha_i$ and $C_i$ label the
center, exponents and contraction coefficients, respectively.
Differentiation of a basis function with respect to these variables leads to another Gaussian function.
As such, derivatives of an ERI can be computed by a sequence of similar integral evaluations.

In the current implementation of {\pyscfad},
the program walks through the computation graph and determines the required intermediate ERIs.
For example, computing the second order derivative
of the one-electron overlap integral with respect to the basis function centers
requires the evaluation of three integrals
\begin{align}
    \nonumber
    \bm{\nabla}_{\mathbf{r}_0} \cdot \bm{\nabla}_{\mathbf{r}_0}\left(\phi_\mu | \phi_\nu \right)
    & \rightarrow \left(\bm{\nabla}_{\mathbf{r}_0} \phi_\mu | \phi_\nu \right) \\ \nonumber
    & \rightarrow
    \begin{cases}
        \left(\bm{\nabla}_{\mathbf{r}_0} \phi_\mu | \bm{\nabla}_{\mathbf{r}_0} \phi_\nu \right) \\
        \left(\bm{\nabla}^2_{\mathbf{r}_0} \phi_\mu | \phi_\nu \right)
    \end{cases} \;,
\end{align}
where in the above, the permutation symmetries have been considered.
In the implementation, these integrals are all evaluated analytically by the highly optimized integral library {\libcint}.\cite{Sun2015}
Thus, the ERIs themselves are treated as elementary functions during the AD procedure, instead of the lower-level arithmetic operations that compute them.

For the commonly used ERIs, up to fourth order derivatives with respect to the
basis function centers are available through {\libcint}. If higher order derivatives are needed, one can
extend {\libcint}
by generating the code to compute the required intermediate integrals before runtime,
with the accompanying automatic code generator.
Only first order derivatives with respect to the exponents and contraction coefficients
have been implemented so far. This should be sufficient for many purposes, such as for basis function optimization.

The procedure above, in a strict sense, is not fully differentiable, because the ERI evaluation is black-box.
However, a general, efficient and fully differentiable implementation for
the ERIs remains challenging. The advent of compiler level AD tools\cite{NEURIPS2020_9332c513}
may facilitate such developments in the future.

\subsection{Density functionals}
A semi-local exchange correlation (XC) density functional can be expressed as
\begin{equation}\label{eq:exc}
    E_{\rm xc} = \int d\mathbf{r} \rho(\mathbf{r}) \varepsilon_{\rm xc} [\rho(\mathbf{r}), \bm{\nabla} \rho(\mathbf{r}), \bm{\nabla}^2 \rho(\mathbf{r}), \tau(\mathbf{r})] \;,
\end{equation}
where $\varepsilon_{\rm xc}$ is the XC energy per particle,
and $\rho$ and $\tau$ label the electron density and the non-interacting kinetic energy density, respectively.
The integration in Eq.~\ref{eq:exc} is usually carried out numerically over a quadrature grid due to the complexity
of XC functionals:
\begin{equation}
    E_{\rm xc} = \sum_i w_i \rho(\mathbf{r}_i) \varepsilon_{\rm xc}(\mathbf{r}_i) \;,
\end{equation}
where $w_i$ is the weight for the $i$-th grid point at position $\mathbf{r}_i$.
The AD of $E_{\rm xc}$ is straightforward
if one uses a fully differentiable implementation of $\varepsilon_{\rm xc}$.
However, we do not pursue that strategy here, since
density derivatives to finite order (usually up to fourth order)
are sufficient in most practical scenarios and these have
already been implemented in efficient density functional libraries such as {\libxc}.\cite{Lehtola2018}
Therefore, we take an approach similar to that introduced above for the AD of ERIs,
where the derivatives of $\varepsilon_{\rm xc}$ with respect to the density variables are computed analytically, while other derivatives ({\eg}, the derivatives of the density variables) are generated by AD.
In particular, {\libxc} provides analytic functional derivatives of $E_{\rm xc}$, from which the derivatives of $\varepsilon_{\rm xc}$ can be easily expressed.
For example, the first order derivative of $\varepsilon_{\rm xc}$ with respect to $\rho$
within the local density approximation (LDA) is obtained as
\begin{equation}
    \frac{\partial \varepsilon_{\rm xc}}{\partial \rho} = \frac{v_{\rm xc}  - \varepsilon_{\rm xc}}{\rho} \;,
\end{equation}
where both $\varepsilon_{\rm xc}$ and $v_{\rm xc}$ (the XC potential)
are computed analytically by {\libxc}.
The higher order derivatives are obtained by recursion, {\eg},
\begin{equation}
    \frac{\partial^2 \varepsilon_{\rm xc}}{\partial \rho^2} = \frac{1}{\rho} \Big(f_{\rm xc} -2\frac{\partial \varepsilon_{\rm xc}}{\partial \rho} \Big) \;,
\end{equation}
where $f_{\rm xc}$ is the XC kernel.

\subsection{Eigendecompositions}
Although {\jax} provides a differentiable implementation of the eigendecomposition,
it does not handle degenerate eigenstates. Usually, in order to determine the derivatives of degenerate
eigenstates, one needs to diagonalize the perturbations of the matrix being decomposed
within the degenerate subspace until the degeneracy is removed.
Such an approach, however, is not well
defined for reverse-mode AD, because the basis in which to expand the degenerate eigenstates is undetermined until
the back propagation is carried out. The development of a general differentiable eigensolver is beyond the
scope of the current work. In the following, we only discuss a workaround for the AD
of eigenvalue problems arising in mean-field calculations.

At the mean-field level, a gauge invariant quantity
will only respond to perturbations that mix
occupied and unoccupied states.
Thus in many cases, the response of degenerate single-particle eigenstates, i.e. orbitals
(which can only be either occupied or unoccupied for gapped systems)
does not contribute to the derivatives, and can be ignored.
Our implementation directly follows the standard analytic formalism derived
from linear response theory.\cite{Pople1979}
For a generalized eigenvalue problem
\begin{equation}
    \mathbf{FC = SC\bm{\varepsilon}} \;,
\end{equation}
where $\bm{\varepsilon}$ and $\mathbf{C}$ denote the eigenvalue and eigenvector matrices, respectively,
their differentials can be expressed as
\begin{equation}
    \partial \varepsilon_{ii} = \left[\mathbf{I\circ (C^\dagger \partial F C
                                                           - C^\dagger \partial S C \bm{\varepsilon})} \right]_{ii} \;,
\end{equation}
and
\begin{equation}\label{eq:mo1}
    \mathbf{\partial C = C \left[W \circ (C^\dagger \partial F C - C^\dagger \partial S C \bm{\varepsilon})
                                                     - I \circ (\frac{1}{2} C^\dagger \partial S C)\right]} \;,
\end{equation}
respectively.
In the two equations above, $\circ$ represents  element-wise multiplication,
\begin{equation}
    I_{ij} =
    \begin{cases}
        1  & \text{if } i = j \text{ or } \varepsilon_i = \varepsilon_j \;, \\
        0 & \text{otherwise} \;,
    \end{cases}
\end{equation}
and
\begin{equation}
    W_{ij} =
    \begin{cases}
        0  & \text{if } i = j \text{ or } \varepsilon_i = \varepsilon_j \;, \\
        \frac{1}{\varepsilon_j - \varepsilon_i} & \text{otherwise} \;.
    \end{cases}
\end{equation}
Similar approaches have also been reported in other works.\cite{Liao2019,Kasim2021}

\subsection{Implicit differentiation of iterative solvers}
Automatic differentiation of optimization problems or iterative solvers is usually done in one of two ways,
namely, unrolling the iterations\cite{Wengert1964,Ablin2020}
or implicit differentiation.\cite{Griewank2008,Andersson2019,Blondel2021}

\begin{figure}
    \fig{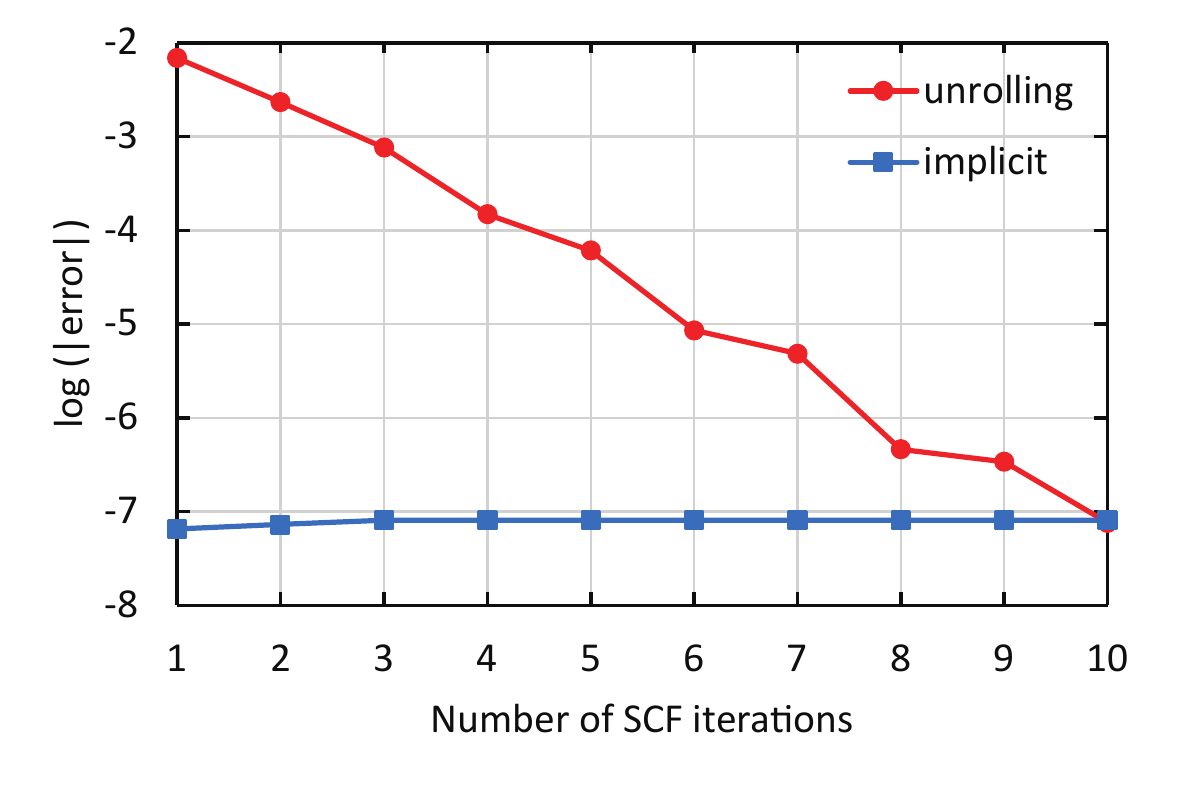}
    \caption{
          Logarithm of the energy nuclear gradient errors plotted as a function of the number of SCF iterations
          for the $\rm N_2$ molecule at the MP2/cc-pVTZ level. The SCF calculations use the converged MOs
          as the initial guess, and two AD approaches were carried out, namely,
          unrolling the iterations and implicit differentiation,
          whose results are shown as the red and blue curves, respectively.
    } \label{fig:mp2_nuc_grad}
\end{figure}

In the first approach, the entire set of iterations is
differentiated, leading to a memory complexity that scales linearly with the number of iterations for reverse-mode
AD. In addition, the so-obtained derivatives are usually initial guess dependent.
This can be understood with the example of
computing the molecular orbital (MO) response (Eq.~\pref{eq:mo1}).
Suppose in an extreme case that
the self-consistent field (SCF) iteration takes the converged MOs as the initial guess,
then the SCF will converge after one Fock matrix diagonalization.
Due to the fact the SCF iteration is short-circuited (the Fock matrix is not rebuilt)
the input MOs have no knowledge of their own derivatives, and
the Fock matrix, which responds to the MO changes, will have the wrong derivative after this single
SCF iteration,
which in turn leads to the wrong MO response.
In other words, self-consistency is not reached when solving Eq.~\pref{eq:mo1},
where $\partial \mathbf{F}$ depends on $\partial \mathbf{C}$;
the quality of convergence of the self-consistent response equations is controlled only by the threshold of the SCF itself.
Such errors can be corrected by increasing the number of SCF iterations,
so that the Fock matrix response with respect to the MO changes is gradually restored, and self-consistency of Eq.~\pref{eq:mo1} is achieved.
To see this effect, we computed the nuclear gradient of N$_2$ molecule
at the MP2 level, where the derivatives of the MO coefficients are
evaluated by the scheme of unrolling the SCF iterations.
Using the same set of converged MOs as the initial guess,
we vary the number of SCF iterations and plot the gradient error,
shown as the red curve in Fig.~\pref{fig:mp2_nuc_grad}.
It is clear that the computed gradient is inaccurate unless a sufficient number of SCF iterations are carried out
to recover the correct MO response. Nevertheless, the method
is problematic when performing AD for optimizations or for iterative solvers,
as one has no direct control over the convergence of the computed derivatives.

In implicit differentiation, instead of differentiating the solver iterations,
the optimality condition is implicitly differentiated. For example,
the solution (denoted as $x^\star$) of the iterative solver should also be the root of some optimality condition
\begin{equation}
    f(x^\star(\theta), \theta) = 0\;.
\end{equation}
According to the implicit function theorem,\cite{Krantz2002,Griewank2008}
$x^\star$ can be seen as an implicit function of $\theta$, and
its derivative can be obtained by solving a set of linear equations
\begin{equation}\label{eq:implicit_diff}
    \frac{\partial f}{\partial x^\star} \frac{\partial x^\star}{\partial \theta} = -\frac{\partial f}{\partial \theta}\;.
\end{equation}
In the problem of computing the MO response, Eq.~\pref{eq:implicit_diff} simply
corresponds to the coupled perturbed Hartree-Fock (CPHF) equations.\cite{Pople1979}
More interestingly, in the reverse-mode AD, one does not directly solve Eq.~\pref{eq:implicit_diff};
instead, the vector Jacobian product
({\ie}, $v^\top J$ where $v$ is a vector and $J \equiv \frac{\partial x^\star}{\partial \theta}$) is computed.
If we define $A \equiv \frac{\partial f}{\partial x^\star}$ and $B \equiv -\frac{\partial f}{\partial \theta}$,
then the vector Jacobian product is obtained as
\begin{equation}\label{eq:vjp}
    v^\top J = z^\top B \;,
\end{equation}
where
\begin{equation} \label{eq:z_vector}
    A^\top z = v \;.
\end{equation}
In this case, only one set of linear equations (Eq.~\pref{eq:z_vector}) needs to be solved
even if derivatives are evaluated with respect to multiple variables
({\ie}, when $B$ changes but not $A$ and $v$).
Note that Eqs.~\pref{eq:vjp} and \pref{eq:z_vector} correspond exactly to the  $Z$-vector approach\cite{Handy1984}
in conventional analytic derivative methods.

The advantages of the implicit differentiation approach are obvious.
First, because only the optimality condition is differentiated,
the computational complexity and memory footprint
do not depend on the actual implementation of the solver or
on the number of solver iterations.
A notable benefit of this is that algorithms to accelerate convergence can be readily applied,
such as the direct inversion of the iterative subspace\cite{Pulay1982} (DIIS) method,
without any modification to the AD implementation.
(Note, however, that Eq.~\pref{eq:z_vector} itself needs to be solved iteratively,
which can be more expensive than the approach of unrolling the iterations,
depending on the size and convergence of the problem).
Second, the computed derivatives have errors that are governed only by the accuracy of the solution
of Eq.~\pref{eq:z_vector}, which can be controlled with a predefined convergence threshold.
This can be seen from the blue curve in Fig.~\pref{fig:mp2_nuc_grad},
where the implicit differentiation approach is applied for
the same MP2 nuclear gradient calculation as discussed above.

Given a general implementation,\cite{Blondel2021}
implicit differentiation
can be applied to almost any iterative solver.
Currently, we have adapted it for the AD of
SCF iterations and the coupled cluster (CC) amplitude equations.
This eliminates the need to explicitly implement and solve
the CPHF equations and the CC $\Lambda$ equations.\cite{Scheiner1987}
Finally, it is interesting to mention that
if the optimality condition being differentiated
involves the fixed point problem of gauge variant quantities ({\eg}, wavefunctions),
it is important to fix the gauge.
In particular, the fixed point of the optimality condition should be identical to the
variable used to evaluate the objective function.

\section{Applications} \label{sec:app}
In this section, we provide a few examples that demonstrate the capabilities of {\pyscfad}.
In addition, complete code snippets are presented to highlight the
ease of developing new methodologies within the framework of {\pyscfad}.
Reverse-mode AD was applied in all the following calculations,
although {\pyscfad} also allows forward-mode AD for its functions.

\subsection{Orbital optimization}
Orbital optimization in electronic structure methods provides the following  advantages:
\begin{enumerate}
    \item Energies become variational with respect to orbital rotations,
    thus there is no need for orbital response when computing nuclear gradients.\cite{Rendell1987}
    \item Properties can be computed more easily because
    there are no orbital response contributions to the density matrices.
    \item The spurious poles in response functions for inexact methods
    such as coupled cluster theory can be removed.\cite{Pedersen2001}
    \item Symmetry breaking problems may be described better.\cite{Sherrill1998}
\end{enumerate}
However, it is not always straightforward to implement analytic orbital gradients
(and hessians for quadratically convergent optimization)
if the underlying electronic structure method is complicated.
In contrast, little effort is needed to obtain the orbital gradients and hessians using AD
as long as the energy can be defined and implemented as differentiable with respect to orbital rotations.

\begin{figure}
    \code{0.8}{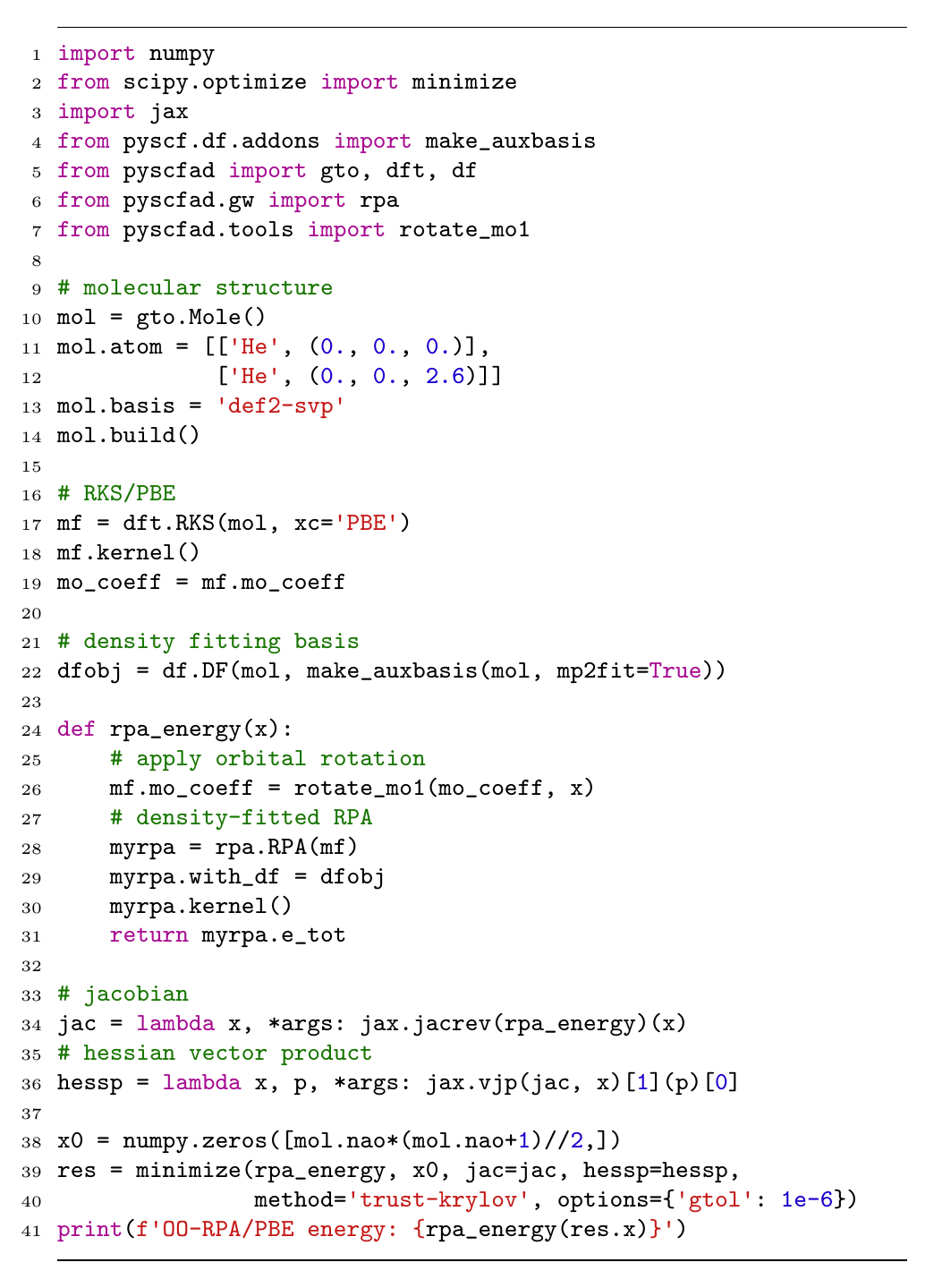}
    \caption{
        Application of orbital optimization for density fitted RPA within the framework of {\pyscfad}.
    } \label{code:oorpa}
\end{figure}

\begin{figure}
    \fig{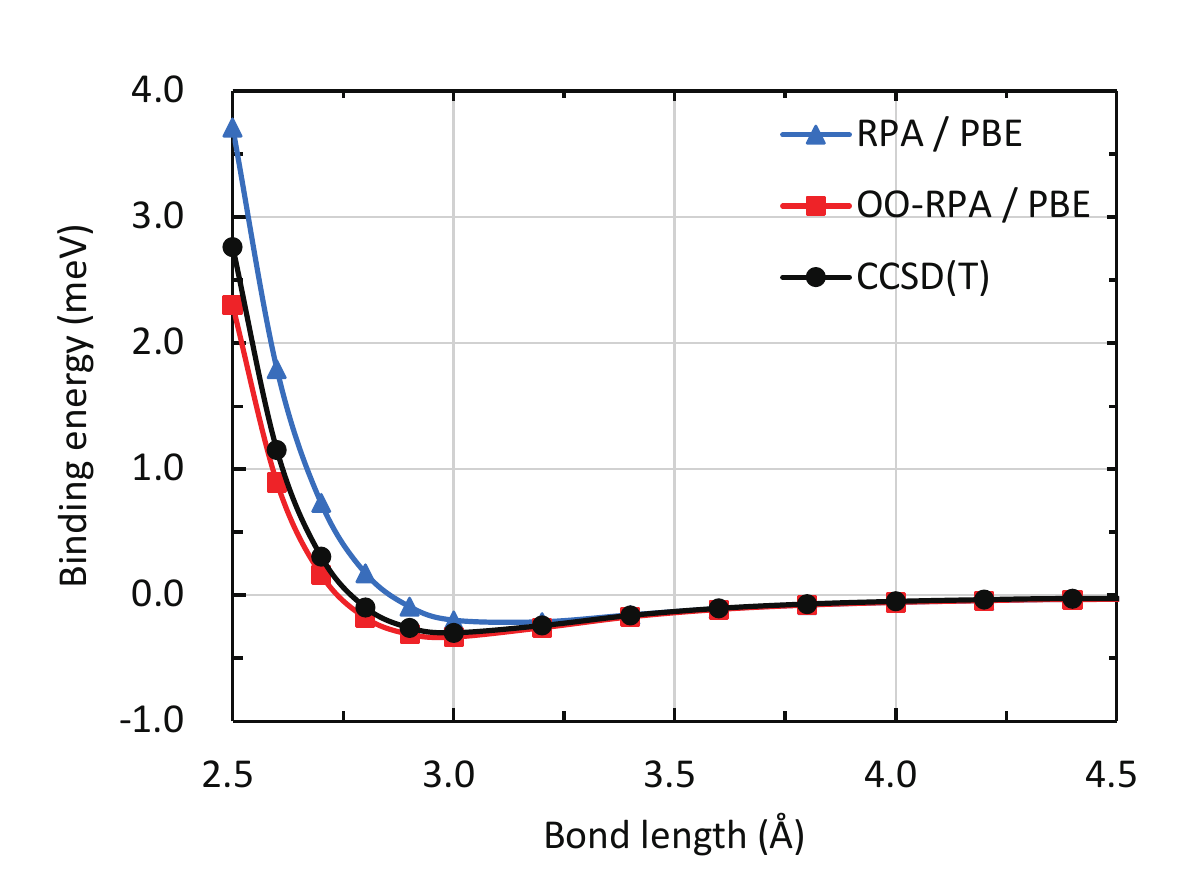}
    \caption{
        Binding energy curves for He$_2$ molecule computed at the RPA, OO-RPA, and CCSD(T) levels.
        The def2-SVP basis set was used throughout and the PBE functional was used for the RPA methods.
        The curves are shifted to match at the bond length of 10.0 {\AA}.
    } \label{fig:rpa_he2}
\end{figure}

As an example, in Fig.~\pref{code:oorpa}, we show the application of orbital optimization to
the random phase approximation for the energy\cite{Bohm1953,Gell-Mann1957,Furche2001} (RPA) within the framework of {\pyscfad}.
The base RPA method is implemented following Ren {\etal},\cite{Ren2012}
where the correlation energy is evaluated by numerical integration over the imaginary frequency axis
and density fitting is applied to reduce the computation cost.
The total energy is then minimized with respect to the orbital rotation $e^\mathbf{x}$,
where $\mathbf{x}$ is anti-hermitian and its upper triangular part
corresponds to the variable ``x'' in Fig.~\pref{code:oorpa}.
Note that only the energy function needs to be explicitly implemented (lines 24--31 in Fig.~\pref{code:oorpa}),
whereas the orbital gradient and hessian (specifically, the hessian-vector product in the example)
are obtained directly by AD (lines 33--36 in Fig.~\pref{code:oorpa}).
The performance of the resulting orbital optimized (OO) RPA method
is shown in Fig.~\pref{fig:rpa_he2}.
We plot the binding energy curves for the He$_2$ molecule computed at the RPA, OO-RPA and
CCSD(T) levels.
It is clear that the molecule is underbound at the RPA level, whereas applying orbital optimization
corrects that and the binding energies are more consistent with the CCSD(T) results.

\subsection{Response properties}
In general, ground-state {\em dynamic} (frequency-dependent) response properties
are related to the derivatives of the quasienergy with respect to  perturbations.\cite{Christiansen1998}
The quasienergy is defined as the time-averaged expectation value of $\hat{H} - i {\partial}/{\partial t}$
over the phase-isolated time-dependent wavefunction.\cite{Christiansen1998}
As such, although there is no difficulty applying AD to compute such derivatives, a straightforward application to the quasienergy in this form requires the
time-dependent wavefunction itself.
In the time-independent limit, however, the quasienergy reduces to the usual energy,
thus the {\em static} response properties can be computed straightforwardly with AD from only time-independent quantities.

\begin{figure}
    \code{0.8}{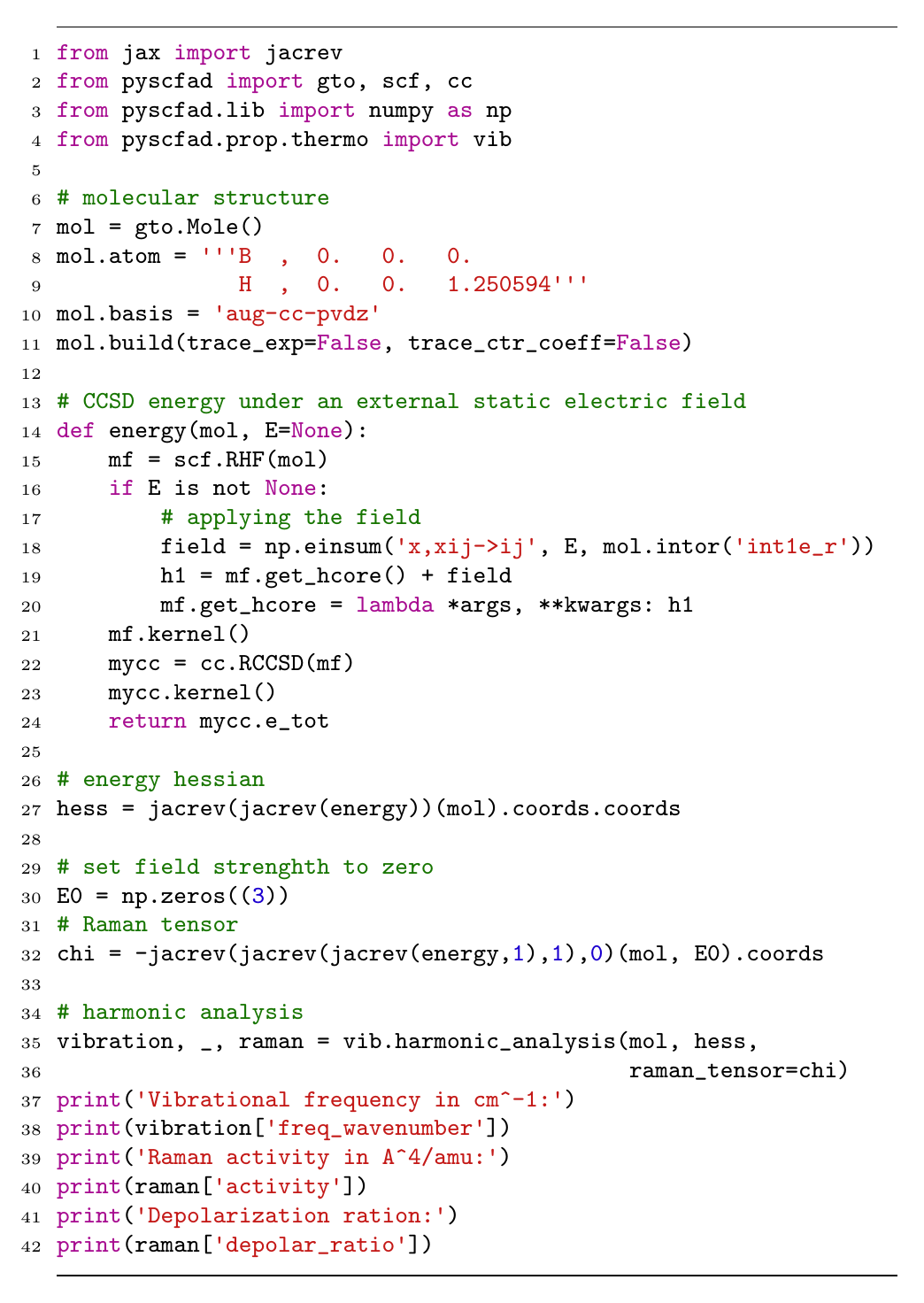}
    \caption{
        An example of computing harmonic vibrational frequencies and Raman activities
        within the framework of {\pyscfad}.
    } \label{code:ccraman}
\end{figure}

For example, the static Raman activity is related to the susceptibility\cite{Placzek1934,ONeill2007}
\begin{equation}
    \bm{\chi} = \frac{\partial^3 E}{\partial \mathbf{R} \partial \bm{\varepsilon}^2} \;,
\end{equation}
where $E$ is the ground-state energy, $\mathbf{R}$ denotes the nuclear coordinates,
and $\bm{\varepsilon}$ represents the electric-field.
In Fig.~\pref{code:ccraman}, we present an implementation of Raman activity at the CCSD level
within the framework of {\pyscfad}.
Again, only the energy function needs to be explicitly implemented (lines 14--24 in Fig.~\pref{code:ccraman}),
while all the derivatives are obtained by AD (lines 27 and 32 in Fig.~\pref{code:ccraman}).
In particular, the CPHF and CC $\Lambda$ equations
are solved implicitly through the implicit differentiation procedure.
Using this code, we compute the harmonic vibrational frequency,
Raman activity and depolarization ratio for the BH molecule,
and the results are displayed in Table~\pref{tb:raman}.
It can be seen that they agree very well with the reference analytic results,
which validates our AD implementation.

Similarly, the IR intensity, which involves computing the nuclear derivative of the dipole moment,
can be obtained in the same manner with AD.
In Table~\pref{tb:ir}, the IR intensities of the H$_2$O molecule at the CCSD level are displayed.
Both AD and analytic evaluation give almost identical results.

\begin{table}
    \centering
    \caption{
        Harmonic vibrational frequency, Raman activity and
        depolarization ratio of the BH molecule computed at the
        CCSD/aug-cc-pVDZ level.
    } \label{tb:raman}
    \begin{tabular}{l@{\hspace{1cm}}r@{\hspace{.7cm}}r}
        \hline\hline
        Properties &{\pyscfad} &Reference$^a$ \\
        \hline
        Frequency (cm$^{-1}$) &2336.70 &2337.24 \\
        Raman activity (\AA$^4/$amu) &217.84 &215.11 \\
        Depolarization ratio &0.56 &0.56 \\
        \hline\hline
    \end{tabular}
    \begin{tablenotes}[flushleft]\fns
        \item $^a$ Reference values obtained from Ref.~\onlinecite{ONeill2007}.
    \end{tablenotes}
\end{table}

\begin{table}
    \centering
    \begin{threeparttable}
        \caption{
            Harmonic vibrational frequencies and IR intensities
            of the H$_2$O molecule computed at the CCSD/cc-pVDZ level.
        }\label{tb:ir}
        \begin{tabular}{l@{\hspace{.5cm}} rr c@{\hspace{.5cm}} rr}
            \hline\hline
            Modes & \mc{2}{c}{Frequency (cm$^{-1}$)} & &  \mc{2}{c}{IR intensity (km/mol)} \\
            \cline{2-3} \cline{5-6}
            & {\pyscfad} & Reference$^a$ & & {\pyscfad} & Reference$^a$ \\
            \hline
            1 &3844 &3846 & &4.47 &4.45 \\
            2 &1700 &1697 & &56.41 &56.15 \\
            3 &3956 &3950 & &22.64 &22.62 \\
            \hline\hline
        \end{tabular}
        \begin{tablenotes}[flushleft]\fns
            \item $^a$ Reference values obtained from the computational chemistry
            comparison and benchmark database.\cite{CCCBDB}
        \end{tablenotes}
    \end{threeparttable}
\end{table}

\subsection{Excitation energies}
\begin{figure}
    \code{0.8}{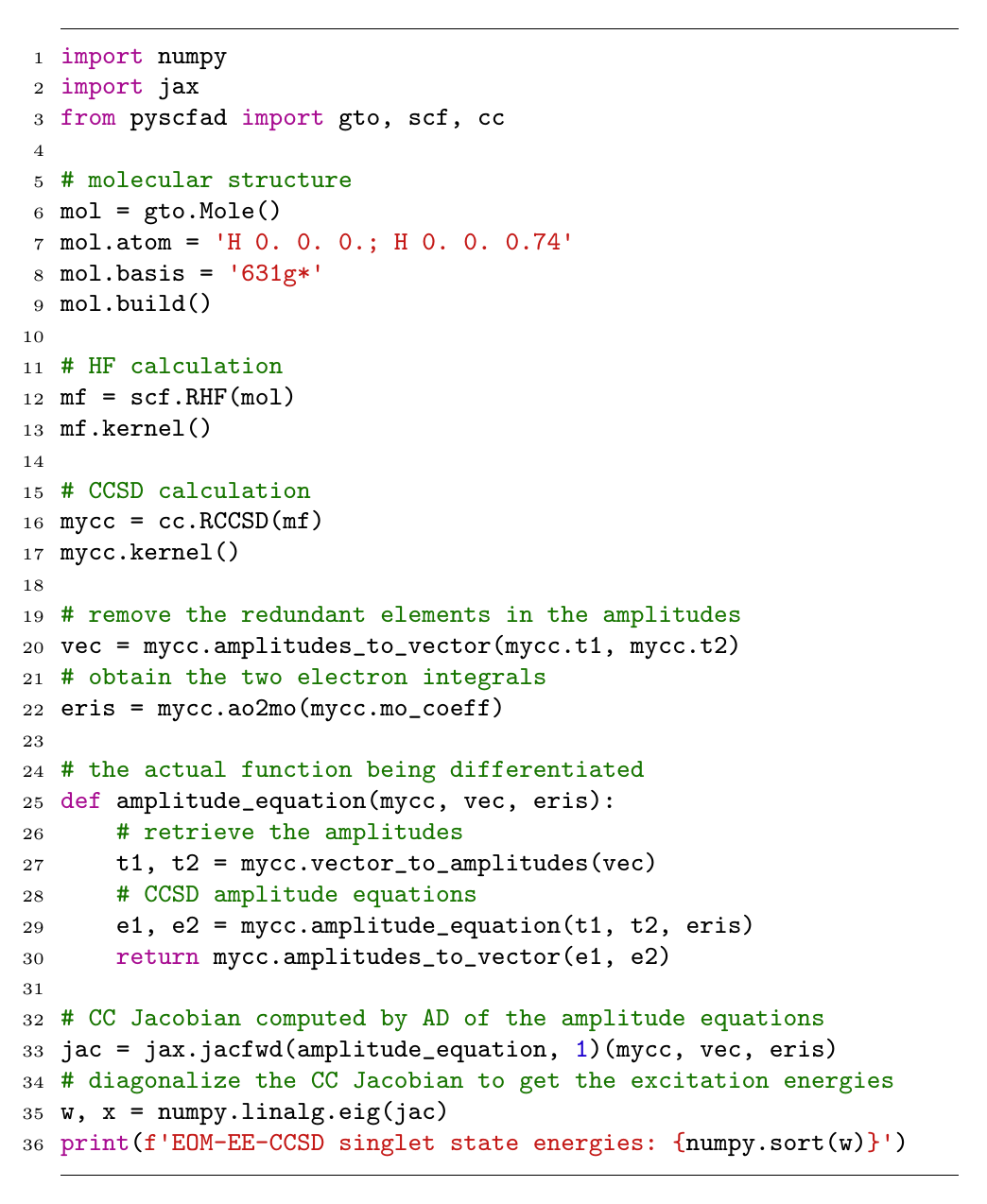}
    \caption{
        An example of computing the CC Jacobian by differentiating the
        amplitude equations within the framework of {\pyscfad}.
    } \label{code:ccjac}
\end{figure}

Excitation energies can be identified from the poles of response functions,
and can be obtained by solving the (generalized) eigenvalue problems which arise from
the derivatives of quasienergy Lagrangians.\cite{Christiansen1998}
For example, in coupled cluster theory,
the excitation energies can be computed as the eigenvalues of the CC Jacobian,
which is defined as the second order derivative of the zeroth order Lagrangian
with respect to the amplitudes and to the multipliers:
\begin{equation}
    \mathbf{A} = \frac{\partial^2 L^{(0)}}{\partial \mathbf{t}^{(0)} \partial \bm{\lambda}^{(0)}} \;.
\end{equation}
Note that $\frac{\partial L^{(0)}}{\partial \bm{\lambda}^{(0)}}$ gives the CC amplitude equations,
which are already implemented in the ground state CC methods.
Therefore, the CC Jacobian and subsequently the excitation energies can be obtained effortlessly
if AD is applied to differentiate the amplitude equations.
In Fig.~\pref{code:ccjac}, we show such an example of computing the excitation energies at the CCSD level,
where only the CC amplitude equations are explicitly implemented (line 29 in Fig.~\pref{code:ccjac}), while the
CC Jacobian is obtained via AD (line 33 Fig.~\pref{code:ccjac}) and is then diagonalized.
The resulting excitation energies are identical to those from the equation-of-motion (EOM) method
(see Table~\pref{tb:eomcc}).
Note that it is also possible to only compute the Jacobian vector product rather than the full Jacobian
(similar to line 36 in Fig.~\pref{code:oorpa}), so that the diagonalization can be performed
with Krylov subspace methods,\cite{Saad2011} making larger calculations practical.

The strategy above is in principle applicable to any method, with the caveat that
complications may arise depending on the particular ansatz used for representing the wavefunction.
In general, the eigenvalue problem being solved has the form
\begin{equation}
    \mathbf{E}^{[2]} \mathbf{x} = \omega \mathbf{S}^{[2]} \mathbf{x} \;,
\end{equation}
 where $\mathbf{E}^{[2]}$ is the second order derivative of the zeroth order energy or Lagrangian,
 and $\mathbf{S}^{[2]}$ may be related to the second order derivative of the wavefunction overlap,
 which may or may not be the identity.

\begin{table}
    \centering
    \caption{
        Excitation energies (in eV) for the lowest four singlet excited states
        of the H$_2$ molecule (with a bond length of 1.1 \AA) computed
        at the EOM-EE-CCSD/6-31G* level.
    }\label{tb:eomcc}
    \begin{tabular}{l@{\hspace{2.8cm}} r@{\hspace{1.5cm}} r@{\hspace{.2cm}} }
        \hline\hline
        States & {\pyscfad}$^a$ & Reference$^b$ \\
        \hline
        S$_1$ & 0.463 & 0.463 \\
        S$_2$ & 0.735 & 0.735 \\
        S$_3$ & 1.096 & 1.096 \\
        S$_4$ & 1.152 & 1.152 \\
        \hline\hline
    \end{tabular}
    \begin{tablenotes}[flushleft]\fns
        \item $^a$ Excitation energies computed by diagonalizing the CC Jacobian from AD.
        \item $^b$ Reference values computed by the EOM-EE-CCSD method.
    \end{tablenotes}
\end{table}

\subsection{Derivative coupling}

The first order derivative coupling
between two electronic states $\Psi_I$ and $\Psi_J$ is defined as\cite{Tully1990,Yarkony1996c}
\begin{equation}
    \mathbf{d}_{IJ} = \langle \Psi_I |\bm{\nabla}_{\mathbf{R}} \Psi_J \rangle \;,
\end{equation}
where $\mathbf{R}$ denotes the nuclear coordinates.
Although derivative couplings are closely related to excited-state nuclear gradients,
their analytic derivation and implementation may still be tedious and error-prone, depending on the
complexity of the underlying electronic structure methods.
However, such calculations can be greatly simplified by applying AD.

\begin{figure}
    \code{0.8}{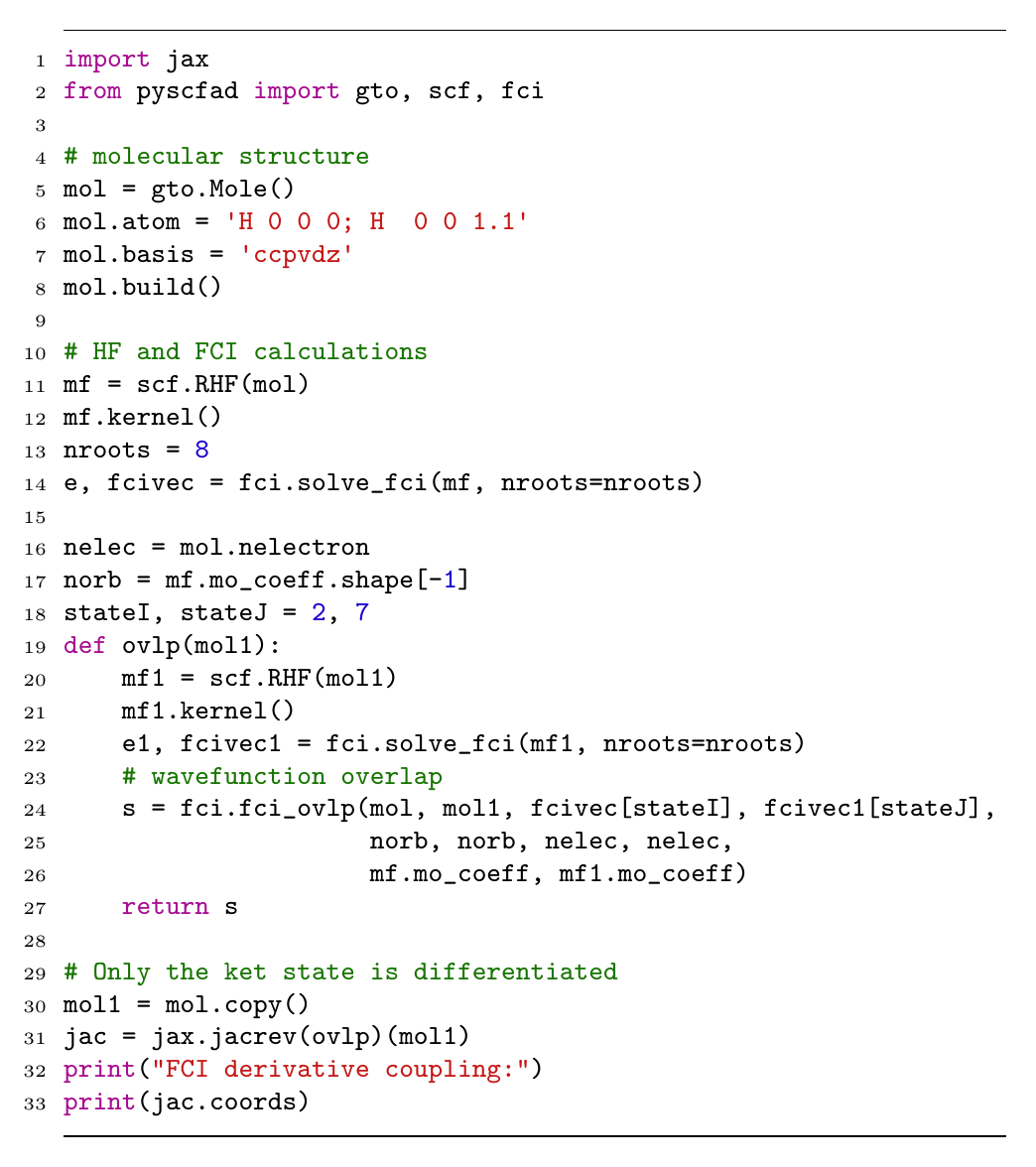}
    \caption{
        An example of computing the derivative coupling between FCI wavefunctions
        within the framework of {\pyscfad}.
    } \label{code:nac}
\end{figure}

In Fig.~\pref{code:nac}, we give an example of computing the derivative coupling at the
full configuration interaction\cite{Knowles1984} (FCI) level
within {\pyscfad}. It should be noted that only a very simple function is
explicitly implemented and then differentiated by AD, which is the
wavefunction overlap $\langle \Psi_I (\mathbf{R}_0) |\Psi_J (\mathbf{R}) \rangle$.
The $\mathbf{R}_0$  here emphasizes that $\Psi_I$ does not respond to the perturbation.
In practice, all the variables corresponding to $\Psi_I$ are held constant when
defining the objective function
({\eg}, ``mol'', ``fcivec'' and ``mf.mo\_coeff'' at lines 24--26 in Fig.~\pref{code:nac}).

The strategy above is applicable to any wavefunction method.
In Table \pref{tb:nac}, we compare the derivative couplings computed at the
configuration interaction singles\cite{Maurice1995} (CIS) and FCI levels.
It is clear that the two methods give very different results.
Nevertheless, our AD implementation produces exactly the same results as the reference implementation.
In addition, second order derivative couplings can be readily obtained by
performing AD on the same wavefunction overlap another time.
Finally, with AD, it is even more straightforward to compute
the Hellmann-Feynman part of the derivative coupling
(which is translationally invariant,\cite{Fatehi2011,Zhang2014a} unlike the full derivative coupling)
\begin{equation}
    \mathbf{h}_{IJ} = \langle \mathbf{C}_I |\bm{\nabla}_{\mathbf{R}} \mathbf{H} | \mathbf{C}_J \rangle \;,
\end{equation}
where $\mathbf{H}$ is the Hamiltonian represented in a certain Hilbert space and
$\mathbf{C}$ denotes the eigenvectors ({\ie}, CI vectors) of $\mathbf{H}$.
Here, only the Hamiltonian needs to be differentiated
(with the orbital response taken into account),
but the CI vectors are treated as constants.
In other words, it is not necessary to differentiate the CI eigensolver
({\eg}, the Davidson iterations\cite{Davidson1975}), which greatly simplifies the problem.

\begin{table}
    \centering
        \caption{
            First order derivative couplings (in a.u.) between the S$_1$ and S$_3$ states
            (both S$_1$ and S$_3$ states have single excitation characters,
            and the S$_3$ state corresponds to the S$_4$ state in FCI calculations)
            for H$_2$ molecule (with a bond length of 1.1 \AA)
            computed at the CIS and FCI levels with the cc-pVDZ basis set.
        }\label{tb:nac}
        \begin{tabular}{l@{\hspace{2.8cm}} r@{\hspace{1.5cm}} r@{\hspace{.2cm}} }
            \hline\hline
            Methods & {\pyscfad} & Reference \\
            \hline
            CIS & 0.0796 & 0.0796$^a$  \\
            FCI & 0.2126 & 0.2126$^b$  \\
            \hline\hline
        \end{tabular}
        \begin{tablenotes}[flushleft]\fns
            \item $^a$ Reference values computed analytically by {\qchem}.\cite{Qchem5}
            \item $^b$ Reference values computed by finite difference.
        \end{tablenotes}
\end{table}

\subsection{Property calculation for solids}
The {\pyscfad} framework also works seamlessly for solid calculations.
Here, we give an example of computing the stress tensor to illustrate this.

The stress tensor $\bm{\sigma}$ is defined as the first order energy response to
the infinitesimal strain deformation\cite{Knuth2015}
\begin{equation}
    \sigma_{\alpha\beta} = \frac{1}{V} \frac{\partial E}{\partial \varepsilon_{\alpha\beta}} \Big |_{\varepsilon=0} \;,
\end{equation}
where $E$ and $V$ are the energy and volume of a unit cell, respectively,
and the strain tensor $\bm{\varepsilon}$
defines the transformation of real space coordinates $\mathbf{R}$ according to
\begin{equation}
    R_\alpha(\bm{\varepsilon}) = \sum_\beta (\delta_{\alpha\beta} + \varepsilon_{\alpha \beta}) R_\beta(\bm{0}) \;,
\end{equation}
in which $\alpha$ and $\beta$ denote the Cartesian components.
The strain deformation applies to both nuclear coordinates and lattice vectors,
which also implicitly affects the derived quantities such as the reciprocal lattice vectors and
the unit cell volume.

\begin{figure}
    \code{0.8}{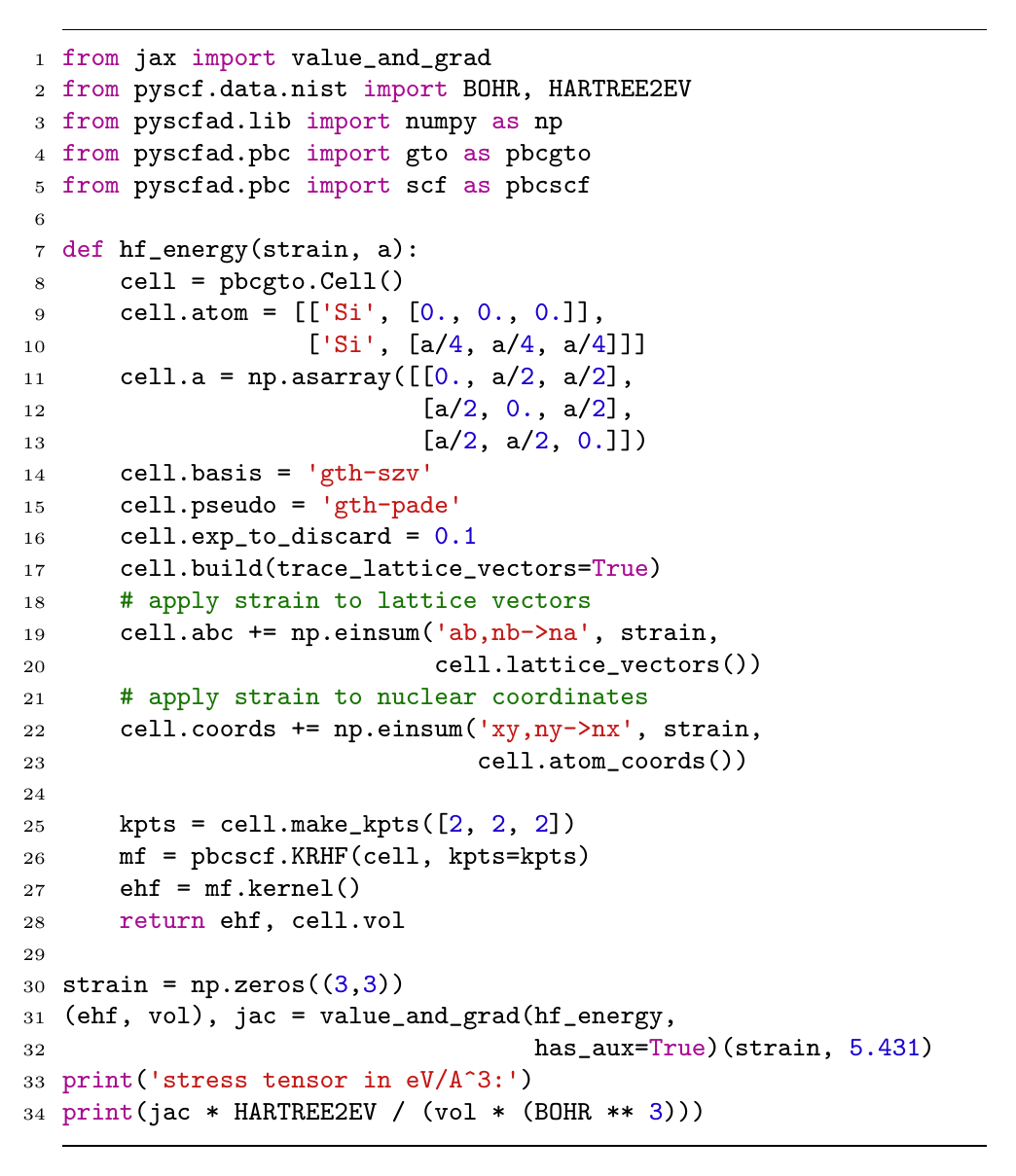}
    \caption{
        An example of computing the stress tensor within the framework of {\pyscfad}.
        The system being studied is a two-atom primitive cell of Si with the lattice parameter of 5.431 \AA.
        The total energy is computed at the HF level using the GTH-SZV basis set
        and the GTH-PADE pseudopotential.\cite{Goedecker1996}
    } \label{code:stress}
\end{figure}

\begin{figure}
    \fig{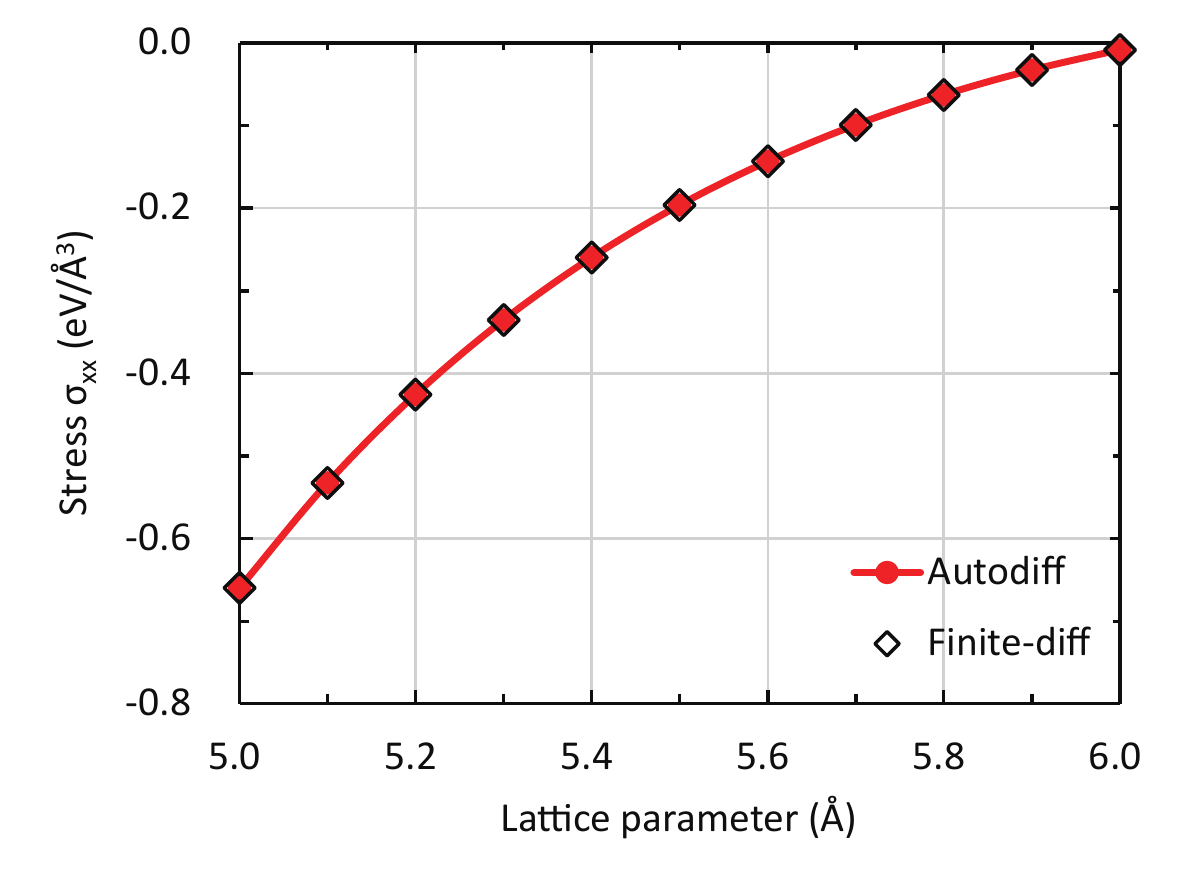}
    \caption{
        Stress tensor component $\sigma_{xx}$ for two-atom Si primitive cells
        computed at the HF level with the GTH-SZV basis and GTH-PADE pseudopotential.
         For various lattice parameters, the results by AD (red) and by finite difference (black) are plotted.
    } \label{fig:stress}
\end{figure}

In Fig.~\pref{code:stress}, the stress tensor for a two-atom Si primitive cell is computed
by AD at the Hartree-Fock level, where a
mixed Gaussian and
plane wave approach (plane-wave density fitting)
is used.\cite{VandeVondele2005,McClain2017}
The results are plotted in Fig.~\pref{fig:stress} along with the corresponding finite difference values.
Perfect agreement is obtained with an average discrepancy of $4 \times 10^{-8}$ eV/\AA$^3$
between the AD and finite difference results.
It should be noted that in this case, the plane wave basis ({\ie}, the uniform grid point) response
to the strain deformation is non-negligible, which subsequently requires
differentiation with respect to the grid points.
However, all these complications are hidden from the user, and it suffices to
modify the energy function (as shown in Fig.~\pref{code:stress}) for the gradient calculations.

\section{Computational efficiency}\label{sec:speed}
In this section, we investigate the computation cost for
carrying out typical quantum chemistry calculations within the {\pyscfad} framework.

\begin{figure}
    \fig{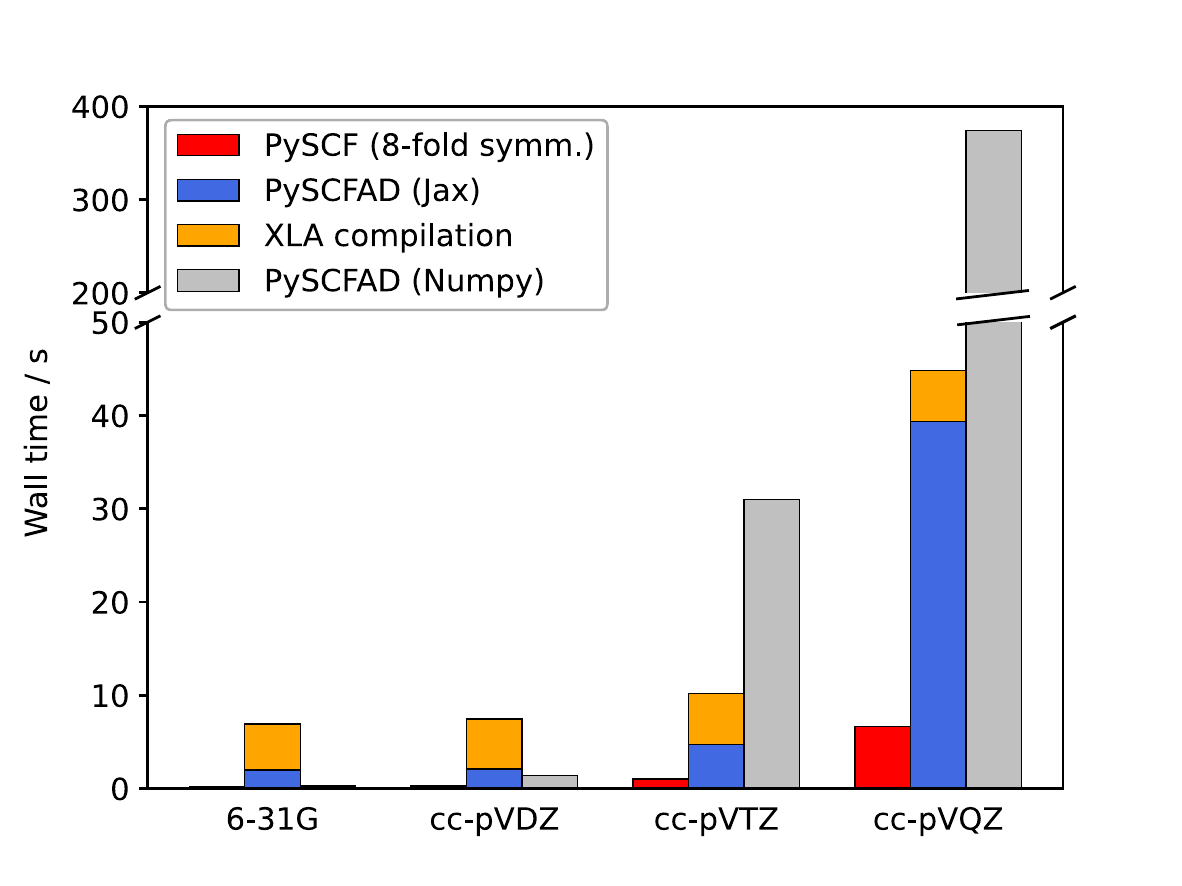}
    \caption{
        Wall time of CCSD(T) energy calculations for the H$_2$O molecule with various basis sets,
        using different implementations. {\pyscf} gives results that are displayed as red bars.
        The blue-and-orange bars represent the total wall time using {\pyscfad},
        where the orange and blue portions correspond to the XLA compilation time and
        the remaining code execution time, respectively.
        The gray bars show the results from the {\numpy} version of the {\pyscfad} implementation
        (where all the {\jax} functions are replaced with their {\numpy} counterparts).
        The calculations were run using 16 Intel Xeon E5-2697 v4 @ 2.30 GHz core.
    } \label{fig:etime}
\end{figure}

First, we consider objective function evaluation.
As an example, in Fig.~\pref{fig:etime},
we plot the computation time for computing the CCSD(T) energy of the H$_2$O
molecule using the implementations in {\pyscf} and {\pyscfad}, respectively.
The main difference between the two implementations is that
{\pyscf} respects the 8-fold permutation symmetry of the
two-electron repulsion integrals, while no such symmetry is enforced in {\pyscfad}.
For {\pyscfad},
we also compare the performance of applying {\jax} and {\numpy} as the
tensor operation backends, respectively.
The corresponding results are presented as the red ({\pyscf}),
blue and orange ({\pyscfad} with {\jax}), and gray ({\pyscfad} with {\numpy}) bars in Fig.~\pref{fig:etime}.
We see that {\pyscfad} with the {\jax} backend is about $5 \sim 8$ times less efficient than {\pyscf},
which is due to the lack of integral symmetries.
However, {\pyscfad} with the {\numpy} backend performs much worse for larger calculations.
This is mainly because the {\em einsum} function in {\numpy}, when applied to
general tensor contractions, may be less optimized and only runs on a single core.
With the {\jax} backend, the same function is optimized by XLA\cite{XLA}
(accelerated linear algebra), which greatly improves the computation efficiency,
although the just-in-time compilation introduces some overhead,
as shown by the orange bars in Fig.~\pref{fig:etime}.
Overall, this suggest that it is potentially possible for {\pyscfad} to be just as efficient as {\pyscf}
for objective function evaluation, if the implementation is carefully optimized.

\begin{figure}
    \fig{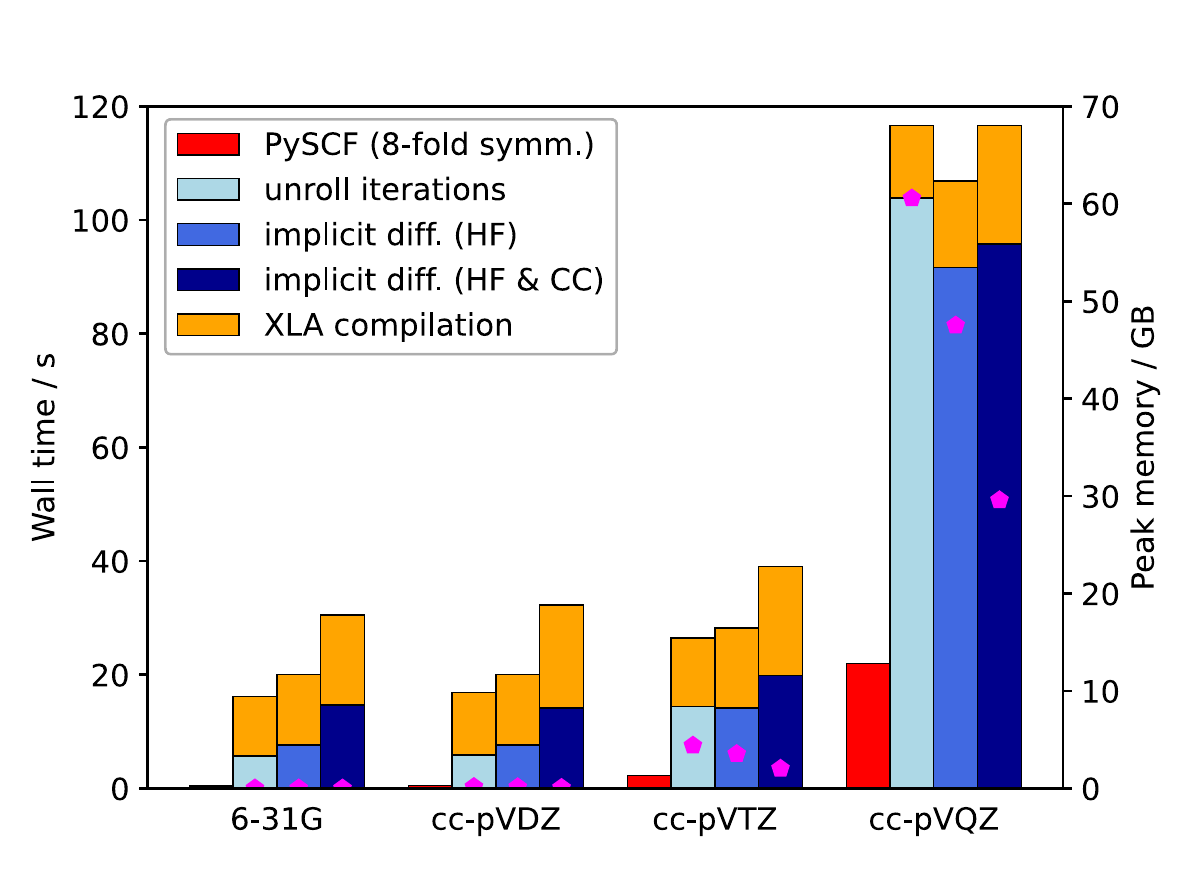}
    \caption{
        Wall time of CCSD nuclear gradient calculations for the H$_2$O molecule with various basis sets,
        using different implementations. {\pyscf} gives results that are displayed as red bars.
        The blue-and-orange bars represent the total wall time using {\pyscfad},
        where the orange and blue portions correspond to the XLA compilation time and
        the remaining code execution time, respectively. Three AD strategies for evaluating the
        derivatives of iterative solvers are investigated,
        namely, unrolling the iterations (light blue), implicit differentiation of the SCF iterations (blue),
        and implicit differentiation of both the SCF iterations and the CC amplitude equations (dark blue).
        The memory footprint of the AD calculations are labeled by the magenta pentagons.
        The calculations were run using 16 Intel Xeon E5-2697 v4 @ 2.30 GHz cores.
    } \label{fig:gtime}
\end{figure}

Second, we examine the efficiency of AD for derivative calculations.
In Fig.~\pref{fig:gtime}, the computation time for CCSD nuclear gradients of H$_2$O
molecule using the analytic implementation in {\pyscf} and the AD procedures in {\pyscfad}
is plotted.
Note that two iterative solvers are involved in the CCSD energy evaluation, {\ie},
the SCF iteration and the CC amplitude equation.
When applying AD to compute the gradient, one can choose to use either the
approach of unrolling the iterations or the implicit differentiation of the two iterative solvers.
As such, the performance of these AD approaches are also compared,
as displayed by the three sets of blue bars in Fig.~\pref{fig:gtime}.
It is clear that all the AD calculations are less efficient than the analytic evaluations by a factor of $5 \sim 8$,
which is again due to the lack of integral symmetries.
On the other hand, the different AD approaches give comparable performance
in terms of the computation time. However, implicit differentiation of the iterative solvers
consumes a considerably smaller amount of memory than unrolling the iterations.
In more detail, for the case of the cc-pVQZ basis set in Fig.~\pref{fig:gtime},
more than 20 gigabytes of memory are used to store the intermediate quantities for computing the
gradients of the ERIs, which applies equally to all the AD approaches.
The remaining memory can be associated with the iterative solvers, and we find that
implicit differentiation of the
SCF and CC iterations reduces the associated memory usage by a factor of $4$ compared to the
approach of unrolling the iterations.
In conclusion, {\pyscfad} appears to be computationally efficient for derivative evaluations,
especially when the iterative solvers are differentiated implicitly.

\section{Conclusions} \label{sec:conc}
In this work, we introduced {\pyscfad}, a differentiable quantum chemistry framework based on {\pyscf}.
It facilitates derivative calculations for complex computational workflows, and can be applied to both molecular and periodic systems, at the mean-field level and beyond.
Using {\pyscfad}, new quantum chemistry methods can be
implemented in a differentiable way with almost no extra effort.
As such, we expect it to become a useful platform to rapidly prototype
new methodologies, which can then be benchmarked on properties
that were previously difficult to compute due to the lack of analytic derivatives.
At the current point, there remain a few challenges to be addressed
in the future developments of {\pyscfad}.
\begin{enumerate}
    \item {\jax} is less efficient than {\numpy} unless the Python functions are compiled with
    XLA. However, this is not always possible, especially when Python control flows
    are involved. As such, code optimization becomes more difficult for {\pyscfad}.
    \item Incorporating permutation symmetries in electron integrals is tricky, because it requires
    element-wise in-place array updates, which are extremely inefficient with current AD tools.
    \item The current implementation amounts to an ``in-core'' version of all methods, {\ie},
    all the data is stored in memory. For ``out-core'' implementations where data can be stored
    on disk, additional work is needed to the track the history of the data,
    without which gradient propagation cannot proceed.
\end{enumerate}

\begin{acknowledgments}
This work was supported by the US Department of Energy through the US DOE, Office of Science, Basic Energy Sciences, Chemical Sciences, Geosciences, and Biosciences Division under Triad National Security, LLC (``Triad'') contract Grant 89233218CNA000001. Support for {\pyscf} ML infrastructure on top of which {\pyscfad} was built comes from the Dreyfus Foundation.
\end{acknowledgments}

\section*{Data availability statement}
The data that support the findings of this study are available from the corresponding author upon reasonable request.
The {\pyscfad} source code can be found at \url{https://github.com/fishjojo/pyscfad}.



\end{document}